# SUMMARY OF THE WORKING GROUP ON 'BEAM–BEAM EXPERIENCE IN HADRON COLLIDERS'

O. Bruning, G. Sterbini, CERN, Geneva, Switzerland

*Abstract*

There were four presentations given during the session 'Beam–beam experience in hadron colliders', reporting the beam–beam observations in SPS [1], Tevatron [2], RHIC [3] and HERA [4]. Presentations were followed by discussions. Below we summarize the major observations, findings and results.

## BEAM–BEAM EFFECTS IN THE SPS PROTON–ANTIPROTON COLLIDER

A rich set of observations and data were collected during the SP**Error! Bookmark not defined.**S proton–antiproton collider run [1].

Without the pretzel separation, only three bunches per beam could be stored in the machine. Using the pretzel separation the number of bunches per beam doubled (6+6). The average beam–beam long-range separation was 3 σ at injection and 6 σ at collision.

The beam–beam effect was a crucial ingredient of the collider beam dynamics. In fact, without separation and with 6+6 bunches, the antiproton beam was lost in less than 5 s.

The tune spread at injection was dominated by space charge in the proton beam and by beam–beam effects in the antiproton beam. The space charge issue was cured with longer bunches in the 200 MHz bucket and by installing a 100 MHz RF system.

During the squeeze it was difficult to keep the tunes constant and the working point of the machine had a strong dependence on the temperature of the final focusing quadrupoles.

From dedicated experiments it was observed that equal geometrical sizes for the colliding beams minimize the beam losses. Colliding beams with different sizes reduce significantly the beam lifetimes and increase the background in the detectors.

Diffusion studies with scrapers during collision showed that the high-order (HO) resonances had almost no effect on the particles with small amplitudes.

Halving the beam separation increased the experiment background by up to a factor 5, whilst no significant effect was observed by varying the crossing angle.

In addition, distance scans were performed to quantify the effect of bunch miss-crossing: it was observed that maximum losses occurred at a separation of 0.2–0.3 σ.

All sources of tune modulation (energy jitter and chromaticity, power supply ripple, etc.) coupled with beam–beam effects had a significant impact and had to be minimized. Using this argument, it was explained that reducing the chromaticity improved the overall machine performance.

*Discussion*

During the discussion it was clarified that, in the past, experiment background was used as the main observable to quantify the beam–beam effect since the detector performance was driven by this figure. Nowadays, with higher peak luminosity achievable, it is no longer a convenient figure of merit and it is replaced mainly by the luminosity lifetime. E. Métral asked if the electrostatic separators used for the pretzel separation are still available at CERN. K. Cornelis answered that this has to be checked, adding that this hardware cannot operate in PPM mode and therefore they are not compatible with the standard SPS supercycle.

## OVERVIEW OF BEAM–BEAM EFFECTS IN THE TEVATRON

The Tevatron has run for more than two decades, constantly pushing its performance beyond the design luminosity [2].

It was already clear in Tevatron Run I that the beam–beam effects would put the beam on dangerous resonances. In fact, already during the 6+6 operation the antiproton emittance blew up; there were halo formation and beam losses, entangling and significant reduction of the luminosity lifetime.

With the installation of the HV separators, helical separation became possible. Orbit and tune variation along the train was observed and it was in very good agreement with simulations.

In Tevatron Run II, the bunches were 36+36. The operations were much more involved. The long-range interaction was critical during injection, ramp and squeeze. The chromaticity played an important role and, as for the SPS, it was important to minimize it. The minimum beam separation at the parasitic beam encounters had to be larger than 5–6 σ.

The total measured HO tune shift went up to 0.04, but the best integrated luminosity could be generated for HO tune shifts around 0.02 and the more dangerous resonances were the 5th, 12th and 7th order. In order to reduce the beam losses, the beam sizes were matched (blow-up of antiproton emittance). A semi-empirical model of integrated luminosity was set up to compute the different contributions to the machine performance: at the end of Run II, the beam–beam effect induced losses accounted for 22–32% of the integrated luminosity.

Several solutions have been adopted over the years to alleviate the beam–beam effect:
- increase the beam separation,
- reduce the Q' and Q'',

- use the transverse damper (at injection) and the octupoles all along the cycle,
- stabilize the orbit and the tune of the machine,
- improve the diagnostics,
- use e-lens compensation.

*Discussion*

During the discussion, there was a question on the limiting factor of the chromaticity correction. A. Valishev answered that the limit was due to the number of sextupole families (driven by the cost of cables). R. Giachino asked to comment on the potential of the Schottky tune monitor. V. Shiltsev answered that it was a valuable tool during operation. O. Bruning asked about the main sources of noise. V. Shiltsev answered that the main contributors were vacuum pumps and bus stability. A. Chao asked about the chromatic effect due to the beam–beam long-range separation. V. Shiltsev answered that was simulated to be ~6 units.

## BEAM–BEAM OBSERVATIONS IN RHIC

The relativistic heavy ion collider (RHIC) at Brookhaven National Laboratory has been in operation since 2000. Over the past decade the luminosity in the polarized proton (p–p) operations has increased by more than one order of magnitude. However, the figure of merit for the p-p operation in RHIC is given by the luminosity times the fourth power of the proton beam polarization. The total peak luminosity therefore plays only a secondary role compared to the beam polarization. The maximum total beam–beam tune shift with two collision points has reached 0.018. The beam–beam interaction leads to large tune spread, emittance growth and short beam and luminosity lifetimes. The longitudinal bunch profiles have large tails due to the re-bunching with a higher RF system and the beam lifetime is therefore sensitive to the non-linear chromaticity.

The main limits to the beam lifetime in the RHIC p-p runs are the beam–beam interaction, the non-linear magnetic field errors in the interaction regions (IRs), the non-linear chromaticities with low-beta, the horizontal and vertical third-order betatron resonances and the machine and beam parameter modulations.

The luminosity decay is fitted with a double exponential: a short and a long time constant were observed (respectively ~1 and ~100 h). Just after the first collisions, a shortening of the longitudinal emittance and a shrinking of the transverse emittance take place systematically, together with beam losses. This effect is higher in bunches with two HO collisions than in bunches with one HO.

From studies and simulation it was concluded that the beam–beam effect reduces the off-momentum transverse dynamic aperture and the fast losses at the start of the physics run are dominated by this mechanism.

During the second part of the run, the present IBS model can fully justify the observed luminosity decay.

To increase the machine performance, chromatic effects of the low-beta played an important role together with the 10 Hz oscillations of the triplets. To attack these issues, several correction techniques of non-linear chromaticities have been tested and implemented in RHIC together with a 10 Hz orbit feedback successfully tested in the 2011 p–p run.

To reduce the large beam–beam tune spread from high bunch intensities, electron lenses are being installed in RHIC.

*Discussion*

A. Valishev asked if local or global chromatic correction was implemented. Y. Luo answered that the correction was local.

## BEAM–BEAM EFFECTS IN HERA

Differently to the three machines previously described, HERA was a hadron–lepton collider (920 GeV p colliding with 27.6 GeV e±) [4].

Due to the filling scheme and to the machine layout there were no long-range encounters, no Pacman bunches, no crossing angle between the beams and the same bunches were crossing at the two Interaction Points, IPs, (one-on-one configuration, no multibunch beam–beam coupling).

Matching the transverse beam sizes at the IPs was mandatory in order to obtain good beam lifetimes (elliptical beams). Due to the induced tune spread, the beam–beam interaction negatively affected the e± polarization.

The choice of the tunes was crucial and the collision tunes for $e^+$, $e^-$ and p were different in order to avoid the second- and third-order sideband resonances and to optimize the lepton polarization.

For HERA beams, partial separation at the IPs was observed to be very problematic.

Strong–strong beam–beam theories do not predict an unstable mode for the HERA parameters. Nevertheless, beam–beam instabilities were observed: this is thought to be due to the driven coherent oscillation on the lepton beam and not to a coherent beam–beam mode.

*Discussion*

A. Burov asked if the damper was used during operation. M. Vogt answered that it was not used for the proton since it was too noisy. W. Fischer asked if there had been attempts to further increase the HO tune shift. M. Vogt replied that there had been few attempts, without positive results.

## CONCLUSIONS

In past hadron colliders it has been observed that:
- The mismatch between the beam sizes can significantly increase the beam–beam detrimental effect.
- Reducing the beam–beam long-range separation below 6 σ had a severe impact on the beam lifetime.

- A crucial way to reduce the beam–beam induced losses is to lower chromaticity and to limit its non-linear component.
- Regarding the HO tune shift limit, it is not possible to extract a general rule since it depends strongly on the observable chosen to define it (experiment background, luminosity lifetime) and on the noise and tune stability of the single machine.

## ACKNOWLEDGEMENTS

The chairman acknowledges all the speakers for the high quality of their presentations.

## REFERENCES

[1] K. Cornelis, "Beam–beam effect in the SPS proton-antiproton collider", these proceedings.
[2] V. Shiltsev, "Overview of beam-beam effect in the Tevatron", these proceedings.
[3] Y. Luo, "Beam–beam observations in RHIC", these proceedings.
[4] M. Vogt, "Beam–beam effects in HERA", these proceedings.